\documentclass[useAMS,usenatbib]{mn2e}
\usepackage{graphicx}

\def\Mo{\ensuremath{M_{\odot}}}

\def\Ro{\ensuremath{R_{\odot}}}
\def\teq{$T_{\rm eq}$}
\def\ergscm{erg\,s$^{-1}$\,cm$^{-2}$}
\def\Mj{\ensuremath{M_{\rm Jup}}}
\def\Rj{\ensuremath{R_{\rm Jup}}}
\def\kms{$\mathrm{km\,s}^{-1}$}

\title[Effect of stellar wind magnetic fields]
{Effect of stellar wind induced magnetic fields on planetary obstacles of non-magnetized hot Jupiters}
\author[N. V. Erkaev  et al.]
{N.~V.~Erkaev$^{1,2}$, P.~Odert$^{3}$, H.~Lammer$^{3}$, K.~G.~Kislyakova$^{4,3}$, L.~Fossati$^3$,\newauthor
 A.~V.~Mezentsev$^{2}$, C.~P. Johnstone$^4$, D.~I.~Kubyshkina$^{3}$, I.~F.~Shaikhislamov$^{5}$,\newauthor M.~L.~Khodachenko$^3$
\\
$^1$Institute of Computational Modelling SB RAS, 660036 Krasnoyarsk, Russia\\
$^2$Siberian Federal University, 660041 Krasnoyarsk, Russia\\
$^3$Space Research Institute, Austrian Academy of Sciences, Schmiedlstrasse 6, A-8042, Graz, Austria\\
$^4$Institute for Astronomy, University of Vienna, T\"{u}rkenschanzstrasse 17, A-1180 Vienna, Austria\\
$^5$Institute of Laser Physics SB RAS, 630090 Novosibirsk, Russia}
\date{}

\pagerange{\pageref{firstpage}--\pageref{lastpage}} \pubyear{2017}

\def\LaTeX{L\kern-.36em\raise.3ex\hbox{a}\kern-.15em
    T\kern-.1667em\lower.7ex\hbox{E}\kern-.125emX}

\begin{document}

\label{firstpage}

\maketitle

\begin{abstract}
We investigate the interaction between the magnetized stellar wind plasma and the partially ionized hydrodynamic hydrogen outflow from the escaping upper atmosphere of non- or weakly magnetized hot Jupiters. We use the well-studied hot Jupiter HD\,209458b as an example for similar exoplanets, assuming a negligible intrinsic magnetic moment. For this planet, the stellar wind plasma interaction forms an obstacle in the planet's upper atmosphere, in which the position of the magnetopause is determined by the condition of pressure balance between the stellar wind and the expanded atmosphere, heated by the stellar extreme ultraviolet (EUV) radiation. We show that the neutral atmospheric atoms penetrate into the region dominated by the stellar wind, where they are ionized by photo-ionization and charge exchange, and then mixed with the stellar wind flow. Using a 3D magnetohydrodynamic (MHD) model, we show that an induced magnetic field forms in front of the planetary obstacle, which appears to be much stronger compared to those produced by the solar wind interaction with Venus and Mars. Depending on the stellar wind parameters, because of the induced magnetic field, the planetary obstacle can move up to $\approx$0.5--1 planetary radii closer to the planet. Finally, we discuss how estimations of the intrinsic magnetic moment of hot Jupiters can be inferred by coupling hydrodynamic upper planetary atmosphere and MHD stellar wind interaction models together with UV observations. In particular, we find that HD\,209458b should likely have an intrinsic magnetic moment of 10--20\% that of Jupiter.
\end{abstract}
\section{INTRODUCTION}
As shown by several studies, the upper atmosphere of hydrogen-dominated exoplanets can develop hydrodynamic outflow conditions if the planet orbits close to the host star and is exposed to sufficiently large extreme ultraviolet (EUV) fluxes \citep{Ye_2004,Tian2005,GarciaMunoz2007,Penz2008,Murray2009,Guo2011,Koskinen2010,Koskinen2013a,Koskinen2013b,Lavv2014,Sha_2014,Khodachenko2015,Ch_2015,Ch_2016,Salz2016,Erkaev2016}. These studies neglected the interaction of the planetary atmosphere with the stellar wind, though in reality the escaping atmospheric particles penetrate into the stellar wind and may exert a strong influence on the wind plasma flow in the vicinity of the planet. None of these studies included also the magnetized stellar wind plasma flow that may interact with the upper atmosphere if the planet has no, or only a weak, intrinsic magnetic moment. The main effect of an intrinsic planetary magnetic field on atmospheric escape is to suppress the outflow and to make it highly anisotropic \citep{Adams2011,Trammell2011,Trammell2014,Owen2014,Khodachenko2015}. However, several studies addressed the interaction between a close-in planet with the host star's wind, but some of them neglected magnetic fields and all just considered a purely hydrodynamic interaction \citep{Stone2009,Bisikalo2013,Tremblin2013,Christie2016}. Other studies, instead, applied MHD models \citep{Cohen2011,Matsakos2015,Tilley2016}, but employing mostly simplified descriptions of the planetary wind.

Recently, \citet{Sha_2016} used a multi-fluid code to study the interaction of a non-magnetized hot Jupiter with the stellar wind, taking into account heating by the stellar extreme ultraviolet (XUV) flux and hydrogen photochemistry to self-consistently model the planetary outflow. However, they did not include the interplanetary magnetic field (IMF) and its effect on the formation of the planetary obstacle, which is the topic of the present study.

Sophisticated MHD simulations have been applied by \citet{Cohen2014} to study the magnetospheric structure of habitable zone planets for different conditions of the stellar wind. Using a multi-species MHD model, \citet{Cohen2015} investigated the stellar wind interaction with a Venus-like demagnetized planet. In addition, a multi-species MHD model, developed previously for Venus, was adapted by \citet{Dong2017} for the calculation of the ion escape process from Proxima Centauri b.

There are many physical processes, thermal and non-thermal, which are responsible for the escape of heavy ions and neutral particles and the importance of accounting for non-thermal escape processes for the solar system planets, and Earth in particular, has been shown for example by \citet{Welling2016}. However, the ion escape caused by the interaction with the stellar wind \citep{Kislyakova2013,Kislyakova2014a,Erkaev2016} and the loss of photochemically produced suprathermal hydrogen atoms \citep{Shematovich2010} from a non- or weakly magnetized HD\,209458b-like hot Jupiter are about an order of magnitude smaller than the thermal escape caused by the absorption of the high-energy stellar flux.

\citet[][hereafter KIS14]{Kislyakova2014b} employed an upper atmosphere-stellar wind interaction particle code that includes acceleration by the stellar radiation pressure, natural and Doppler spectral line broadening, and charge exchange with the stellar wind to reproduce the Hubble Space Telescope (HST) Ly-$\alpha$ transit observations of HD\,209458b \citep{Vidal-Madjar2003,Be_2007}. The best fit to the observed Ly-$\alpha$ absorption was obtained for a planetary magnetic field smaller than 0.4\,G.  The results of KIS14 do not support a magnetic moment much larger than about 10\% of Jupiter's, in agreement with previous studies related to the non-detection of exoplanetary radio emission from hot Jupiters, which suggested that because of tidal locking, hot Jupiters may have weak magnetic moments \citep{Griessmeier2004,Griessmeier2007,Weber2017}. As shown by \citet{Khodachenko2015}, such weak intrinsic magnetic fields do not significantly influence the atmospheric outflow.

Following the indirect evidence that HD\,209458b may have a weak intrinsic magnetic field, we investigate the build-up of a planetary obstacle produced by the interaction of the partially ionized planetary wind with the plasma flow of a magnetized stellar wind. In Sect.~\ref{sec:description}, we describe the input parameters and the adopted modeling scheme. In Sect.~\ref{sec:results}, we present our results and discuss the influence of the assumed stellar wind plasma parameters on the obstacle formation and the possible implications for UV observations. Finally, we gather our conclusions in Sect.~\ref{sec:conclusion}.
\section{MODEL DESCRIPTION}\label{sec:description}
\subsection{3D MHD Flow Model}
We use a 3D MHD flow model, based on the scheme of \citet{Far2008,Far2009}, to compute the plasma flow around a non-magnetized HD\,209458b-like planet and to calculate the steady-state plasma parameters and magnetic field in the environment surrounding the planet, considering different stellar wind conditions. This model allows us to calculate the induced electric currents due to the ionization and charge-exchange processes acting on the hydrodynamically expanding upper planetary atmosphere. Such currents produce an induced magnetic field, which can strongly affect the location of the boundary of the planetary obstacle.

The model solves the following equations
\begin{eqnarray}
\frac{\partial (\rho \bf V)}{\partial t}  + \nabla \cdot \left[ \rho {\bf V} {\bf V} + {\bf I}
\left (P + \frac{B^2}{8\pi}\right) -\frac{{\bf B}{\bf B}}{4\pi} \right ] = \nonumber \\
 = Q_i {\bf V_h}  - Q _{ex}({\bf V} - {\bf V}_h), \\
\nabla\cdot {\bf B} =0, \\
\frac{\partial \rho}{\partial t}  + \nabla \cdot (\rho {\bf V}) = Q_i, \\
\frac{\partial \bf B}{\partial t} - \nabla \times ({\bf V} \times {\bf B}) =0, \\
\frac{\partial {W}}{\partial t}  + \nabla \cdot \left(\frac{1}{2}\rho V^2 {\bf V} +
\frac{\gamma}{\gamma-1} P {\bf V} + \frac{1}{4\pi}{\bf B}\times ({\bf V}\times {\bf B})  \right) = \nonumber \\
\left(Q_i + Q_{ex}\right)\left(\frac{1}{2}V_h^2 + \frac{3 k T_h}{2 m_p} \right)  - Q_{ex}\left(\frac{1}{2}V^2 + \frac{3 k T}{2 m_p} \right), \\
W = \frac{\rho}{2} V^2 + \frac{1}{\gamma -1} P  + \frac{1}{8\pi} B^2,
\end{eqnarray}
where $\rho$, ${\bf V}$, $P$, and ${\bf B}$ are the mass density, velocity, plasma pressure, and magnetic field of the stellar wind, respectively. The parameter $\gamma$ is the polytropic index (assumed to be equal to 5/3), while $V_h$ and $T_h$ are the velocity and temperature of the escaping atmospheric neutral hydrogen atoms.

The mass conservation equation includes an interaction source term, which is related to photoionization
\begin{equation}
Q_i = \alpha_i  N_{hn} m_p
\end{equation}
and charge exchange ionization
\begin{equation}
Q_{ex} = \rho <V_{rel}> N_{hn} \sigma_{ex}
\end{equation}
of the hydrogen atoms. Here, $N_{hn}$ is the number density of the neutral planetary hydrogen atoms, $m_p$ the particle mass, $\sigma_{ex}$ ($\sim$10$^{-15}$\,cm$^2$) the charge exchange cross section, $<V_{rel}>$ the average relative speed of the stellar wind and atmospheric particles, and $\alpha$ is the ionization rate proportional to the EUV flux $\alpha$ = 5.9$\times$10$^{-8}$ $I_{EUV}$ \,s$^{-1}$ .

We apply a Godunov-type finite difference method for the numerical calculations of the non-stationary MHD flow around the planetary obstacle, which can briefly be described by rewriting the system of equations in a compact vector form as
\begin{equation}
\frac{\partial U}{\partial t} + \frac{1}{r^2} \frac{\partial (r^2 \Gamma)}{\partial r} +
\frac{1}{r\sin(\theta)}\frac{\partial (\sin(\theta) \Psi)}{\partial \theta}
 + \frac{1}{r\sin(\theta)} \frac{\partial \Phi}{\partial \phi} = G,
\end{equation}
where the vector quantity $U$ consists of the mass, momentum, energy densities, and magnetic field components. The quantities $\Gamma$, $\Psi$, and $\Phi$ denote instead the corresponding fluxes in the radial, meridional, and azimuthal directions, respectively, and the vector $G$ consists of the corresponding source terms.

We employ a finite difference approximation of the MHD equations, which is based on a conservative Godunov-type scheme,
\begin{eqnarray}
\frac{\left (U_{i,j,k}^{n+1} - U_{i,j,k}^{n}\right)}{\Delta t} +
\frac{1}{r_{i}^2} \frac{\left ( r_{i+1/2}^2 \Gamma_{i+1/2,j,k}^{n+1/2}-r_{i-1/2}^2 \Gamma_{i-1/2,j,k}^{n+1/2} \right)}{ \Delta r} \nonumber \\
+ \frac{1}{r_i \sin(\theta_{j})}\frac{\left( \sin(\theta_{j+1/2}) \Psi_{i,j+1/2,k}^{n+1/2} - \sin(\theta_{j-1/2}) \Psi_{i,j-1/2,k}^{n+1/2}    \right)}{\Delta \theta}  \nonumber \\
+ \frac{1}{r_i\sin(\theta_j)} \frac{\left  (\Phi_{i,j,k+1/2}^{n+1/2}- \Phi_{i,j,k-1/2}^{n+1/2}  \right )}{\Delta \phi} = G_{i,j,k}^n,
\end{eqnarray}
where the quantities with half-integer numbers correspond to intermediate time steps. These quantities are determined on the basis of the approximate Riemann solver. A small region around the polar axis ($\theta=0$) is treated separately in a local Cartesian coordinate system in order to avoid singularities.  The condition $\nabla\cdot B=0$ is maintained employing the method proposed by \citet{Powell1994} and \citet{Powell1999}, with the modifications of \citet{Janh2000}.

The stellar wind flow is loaded by newly born planetary ions generated by charge exchange with the exoplanet's neutral exosphere. The mass loading process results in a strong deceleration of the stellar wind plasma and in an enhancement of the magnetic field in front of the obstacle, which corresponds to the formation of an induced magnetosphere. We approximate the streamlined obstacle by a semi-sphere. The distance between the planet and the stagnation point of the stellar wind ($R_s$) is determined by the pressure balance condition. This means that the total pressure of the external magnetized stellar wind flow at the stagnation point has to be equal to the sum of the thermal and dynamic pressures of the internal atmospheric flow.

We note that we shift the planetary center by some distance $d$ towards the star. We do this to avoid the violation of the pressure balance at the flanks, where the stellar wind pressure at the obstacle boundary decreases substantially. We assume a $d/R_s$ distance ratio of 0.3. The calculation domain for the MHD stellar wind flow is bound by the external semi-sphere related to the undisturbed stellar wind region and the internal semi-sphere corresponding to the streamlined obstacle. At the outer boundary, we set the undisturbed stellar wind parameters: density, velocity, temperature, and magnetic field. At the obstacle boundary, we set zero conditions for the normal components of the stellar wind velocity and magnetic field. We obtain a steady-state solution as a result of time relaxation of the non-steady MHD solution. As initial conditions, we apply the undisturbed stellar wind parameters in the computational domain.
\subsection{Hydrodynamic Upper Atmosphere Model}
To study the EUV heating and expansion and to infer the mass-loss rates of the hydrogen-dominated upper atmosphere of the planet, we apply a time-dependent 1D hydrodynamic model described in detail by \citet{Erkaev2016}. The model solves the absorption of the stellar EUV flux by the thermosphere, the hydrodynamic equations for mass, momentum, and energy conservation, and the continuity equations for neutrals and ions (both atoms and molecules). The code also accounts for dissociation, ionization, recombination, and Ly-$\alpha$ cooling. The quasi-neutrality condition determines the electron density.

The model does not self-consistently calculate the ratio of the net local heating rate to the rate of the stellar radiative energy absorption. In general, this ratio, also called heating efficiency, is not constant with altitude. Studies solving the kinetic Boltzmann equation applying a direct-simulation Monte Carlo model to calculate the heating efficiency indicate values between 10 and 20\% \citep{She_2014}. We therefore adopt a heating efficiency of 15\%, which is in good agreement with what obtained by \citet{Owen2012}, \citet{She_2014}, and \citet{Salz2016}.

As in \citet{Murray2009}, we assume a single wavelength for all EUV photons ($h\nu$\,=\,20\,eV) and use an average EUV photoabsorption cross section for hydrogen atoms and molecules of 2$\times$10$^{-18}$\,cm$^{2}$ and 1.2$\times$10$^{-18}$\,cm$^{2}$, respectively, which are in agreement with experimental data and theoretical calculations \citep[e.g.][]{Be_1964}.

In the framework of our spherically symmetric hydrodynamic model for the planetary atmosphere, we can describe only radial variations of the atmospheric density, pressure, velocity, and temperature. Since we are not capable to describe angular variations of the atmospheric parameters (which requires a 2D model), we can fulfill the pressure balance condition just at the central stagnation point, where the stellar wind velocity goes to zero. By applying this condition, we determine the radial distance between the stagnation point and the planetary center. In particular, we use the profiles for the atmospheric parameters and find the point where the sum of atmospheric thermal and dynamic pressures is equal to the stellar wind total pressure. The latter is the sum of the magnetic and plasma pressures, which depend on the stellar wind upstream input parameters.
\subsection{Planetary and Stellar Input Parameters}
For the stellar and planetary system parameters (i.e., planetary mass $M_{\rm p}$, planetary radius $R_{\rm p}$, equilibrium temperature $T_{\rm eq}$, orbital separation $a$, stellar mass $M_{\rm star}$, and stellar radius $R_{\rm star}$) we adopt the values of HD\,209458 and HD\,209458b from \citet{Southworth2010}. Following the results of \citet{Lam2016}, \citet{Cub_2017}, and \citet{fossati2017}, we fix the lower boundary for the hydrodynamic model of the planetary upper atmosphere at the optical transit radius. Here, we assume that the pressure ($P_0$) is equal to 100\,mbar and the temperature is equal to the equilibrium temperature (\teq). For the stellar EUV flux at the planet's orbit ($I_{\rm EUV}$), we adopt the value given by \citet{guo2016}.

In this work we consider both a slow and a fast stellar wind. The host star HD\,209458 is similar to the Sun, both in terms of mass and age. Since observations cannot directly constrain the stellar wind parameters, we consider two sets of parameters obtained from the solar wind models presented by \citet{Johnstone2015}, plus those inferred by KIS14 derived from fitting the HST Ly-$\alpha$ transit observations. For the latter scenario, we study cases with and without a stellar magnetic field. The stellar magnetic field value was estimated by rescaling the solar interplanetary magnetic field at the Earth's orbit. The full set of adopted input parameters is given in Table~\ref{tab.input_params}.

\begin{table}
	\caption{Input parameters of the simulations. The stellar wind parameters, denoted by the index $w$, correspond to the values at the planetary orbit of 0.047 AU.  Case 1 corresponds to a slow and Case 2 to a fast stellar wind with the inclination angle between the IMF and plasma velocity directions in the undisturbed stellar wind of $\theta_{\rm B}$\,=\,90$\degr$. Case 3 corresponds to the slow wind, but assumes a $\theta_{\rm B}$\,=\,45$\degr$. Cases 4 and 5 correspond to the stellar wind parameters obtained by KIS14, where the former neglects $B_{\rm w}$ and the latter assumes $B_{\rm w}$\,=\,0.014\,G.}
	\label{tab.input_params}
	\begin{center}
\fontsize{6}{6}\selectfont
		\begin{tabular}{llllll}
			\hline
			Stellar  & Case 1 & Case 2&  Case3 & Case 4 & Case 5  \\
            wind   &        &     &    &      &   \\
            			\hline
			$N_{\rm w} $[cm$^{-3}$]   &  4045     &    1371  & 4045    &  5000 & 5000\\
			$V_{\rm w}$[\kms]         &  236    &   532     &236    &   400 & 400\\
			$T_{\rm w} $ [10$^6$ K]    & 1.3   & 2.9  & 1.3        &  1.1  & 1.1 \\
			$B_{\rm w}$ [G]             & 0.014 & 0.014   & 0.014  & 0.0 & 0.014 \\
             $ \theta_{\rm B}$ [\degr]    & 90 & 90 & 45 &   --- &   90  \\
			\hline
			Planetary  &&&&& \\
parameters    &&&&&\\
			\hline
			$M_{\rm p}$\  [\Mj] & \multicolumn{5}{c}{0.714} \\
			$R_{\rm p}$\  [\Rj] & \multicolumn{5}{c}{1.380} \\
			\teq\  [K]   & \multicolumn{5}{c}{1459} \\
			$P_0$  [bar] & \multicolumn{5}{c}{0.1} \\
			$a$ [AU] & \multicolumn{5}{c}{0.047} \\
			\hline
			Stellar &&&&& \\
parameters &&&&&\\
			\hline
			$M_{\rm star}$\  [\Mo]             & \multicolumn{5}{c}{1.148} \\
			$R_{\rm star}$\  [\Ro]             & \multicolumn{5}{c}{1.162} \\
			$I_{\rm EUV}$ [\ergscm] & \multicolumn{5}{c}{1086}\\
			\hline
		\end{tabular}
\end{center}
\textit{Note.} The parameters $M_{\rm p}$ and $R_{\rm p}$ are the planetary mass and radius, respectively, while $M_{\rm star}$ and $R_{\rm star}$ are those of the host star, and $a$ is the orbital semi-major axis.
\end{table}
\section{RESULTS AND DISCUSSION}\label{sec:results}
Assuming that HD\,209458b has a negligible intrinsic magnetic field, and by considering the input parameters given in Table\,1, we employ our 1D and 3D models to obtain the induced magnetic field strength and the stellar wind plasma parameters of the flow around HD\,209458b's planetary obstacle.

Figure~1 shows the spatial distribution of the magnetic field strength obtained from the numerical MHD model employing the Case 1 stellar wind parameters (slow wind). Here, the magnetic field is given in units of the stellar wind magnetic field, $B_{\rm w}$, which is equal to 0.014\,G. The origin of the coordinate system is at planet center and the star is located along the $X$-axis, while the $Z$-axis is directed along the direction of $B_{\rm w}$ (the arrow in Fig.~1), which is assumed to be perpendicular to the  undisturbed wind velocity (in the reference frame of the planet). The white area close to the origin of the $Z$-axis indicates a region filled just by atmospheric particles, while the dark blue semi-circle indicates the planetary surface. The magnetic field has strong pile up in front of the stagnation point. In this region of strong induced magnetic field the magnetic pressure exceeds the local thermal gas pressure of the stellar wind loaded by the outflowing atmospheric atoms.
\begin{figure}
\centering\includegraphics[width=\hsize,clip]{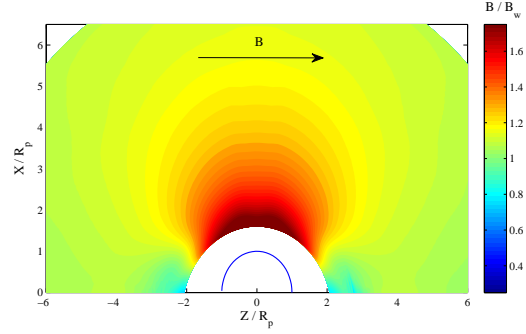}
\caption{Cut at $Y$\,=\,0 of the simulation showing the distribution of the magnetic field strength around HD\,209458b normalized to the IMF of 0.014\,G for a slow stellar wind (Case\,1). The white area close to the origin of the $Z$-axis indicates the atmospheric region around the planet, while the half-circle indicates the planetary optical radius. The star is located along the $X$-axis. The arrow shows the direction of the interplanetary magnetic field.}
\end{figure}
\begin{figure}
\centering\includegraphics[width=\hsize,clip]{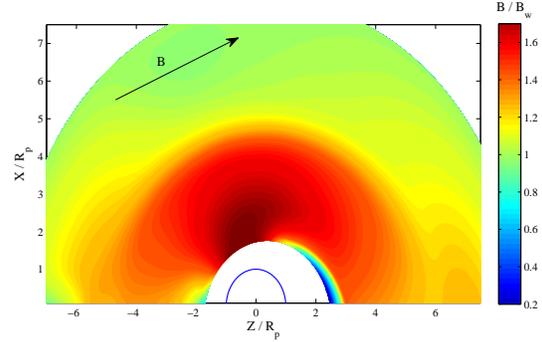}
\caption{Same as Fig.~1, but for an IMF inclined by $45\degr$ (Case\,3).}
\end{figure}
To estimate the influence of the IMF rotation angle, we consider in Case 3 an inclination angle of $\theta_{\rm B}=45\degr$ between the IMF and plasma velocity directions in the undisturbed stellar wind. The calculated distribution of the magnetic field intensity is shown in Fig.~2. One can see that the inclination angle of the magnetic field leads to an asymmetry of the flow structure and of the magnetopause position with respect to the planet. The maximum of the total pressure is also shifted away from the $X$-axis.
\begin{figure}
\centering\includegraphics[width=\hsize,clip]{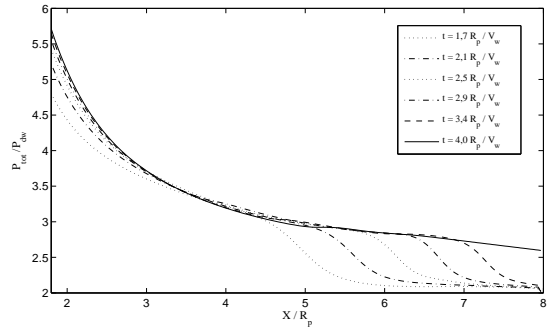}
\caption{Radial profiles along the line connecting the planetary center and the stagnation point for the total pressure (sum of the magnetic and thermal pressures) at six different calculation times.}
\end{figure}

By solving the non-steady MHD equations, a stationary flow pattern is formed after some relaxation period. Initially, we assume that the uniform stellar wind flow is suddenly stopped at the planetary obstacle. This results in the appearance of a shock-like front close to the obstacle and which propagates outwards from it. Since we have a sub-Alfvenic stellar wind flow, this shock propagates far away towards the star. The time dependent propagation of the shock is indicated in Fig.~3, which shows the behaviour of the total pressure along the line connecting the planetary center and the stagnation point (hereafter stagnation line) at different calculation times. We reach a stationary profile after a time corresponding to about 4$\times R_p/V_w$.

Figure~4 shows the magnetic field strength, the total stellar wind pressure (sum of magnetic and thermal pressures), and the thermal stellar wind pressure as a function of the radial distance along the stagnation line for the stellar wind parameters of Case\,1 (solid line) and Case\,2 (dashed line). The magnetic field is given in units of IMF in the undisturbed stellar wind. The magnetic field and total pressure experience a substantial enhancement along the stagnation line from the star towards the planet and reaches its maximum at the stagnation point.
The magnetic field at the stagnation point is increased by a factor of $\la2$ compared to the interplanetary value. The thermal pressure has instead an opposite behavior compared to the total pressure as it decreases to small values at the stagnation point.
\begin{figure}
\centering\includegraphics[width=\hsize,clip]{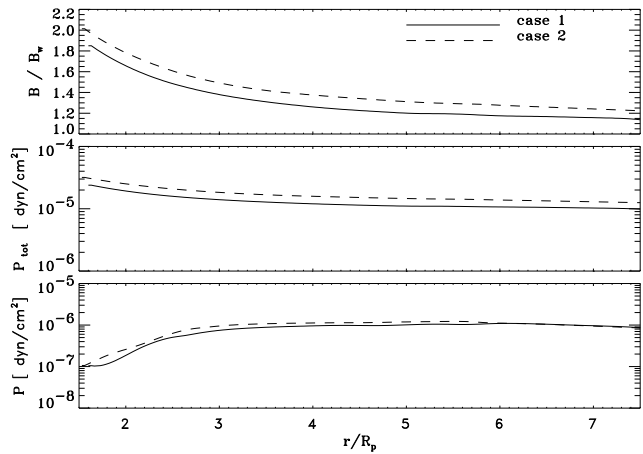}
\caption{From top to bottom: radial profiles along the stagnation line for the magnetic field strength, total pressure (sum of the magnetic and thermal pressures), and thermal pressure. The solid and dashed lines correspond to the Cases 1 and 2, respectively. The planet center is located at $r$\,=\,0.}
\end{figure}

Figure~5 shows the velocity and density of the stellar wind plasma as a function of radial distance along the stagnation line. The total ion number density further increases and has a pronounced maximum of about 4$\times$10$^4$\,cm$^{-3}$ at the stagnation point. This is due to the loading of the stellar wind plasma by newly ionized particles penetrating into the flow from the planet's upper atmosphere. The strong deceleration of the stellar wind plasma near the stagnation point is caused by the appearance of the newly ionized slow particles from the upper planetary atmosphere, which are created via charge exchange and immediately mixed into the plasma flow environment.
\begin{figure}
\centering\includegraphics[width=\hsize,clip]{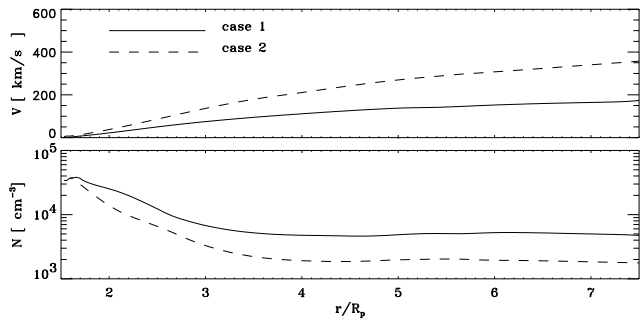}
\caption{Stellar wind velocity and number density as a function of the radial distance along the stagnation line for Case 1 (solid line) and Case 2 (dashed line). The planet is located at $r$\,=\,0. }
\end{figure}

Figure 6 shows HD 209458b's radial profiles of the upper atmosphere's hydrodynamic parameters. The top panel shows the radial profiles of the planetary upper atmospheric number densities of molecular hydrogen, total atomic hydrogen (neutrals and ions), and electrons. The middle panel shows the temperature and velocity of the escaping atmospheric particles as a function of the radial distance. The bottom panel presents the sum of the dynamic and thermal pressures ($\Pi$) in the atmospheric hydrodynamic flow as a function of the radial distance. The diamonds and vertical dotted lines indicate pressure balance distances between the stellar wind plasma and the planetary outflow, corresponding to the stellar wind parameters of cases 1 to 5 (Table 1). One can see, that in cases 1 (slow wind) and 2 (fast wind) with $ \theta_{\rm B}=90\degr$ the pressure balance distances are rather close to each other. Comparing the slow wind cases, Case 1 (with IMF and $\theta_{\rm B}=90\degr$) with Case 3 (with IMF and $\theta_{\rm B}=45\degr$), one can conclude that the angular difference of $90\degr \rightarrow 45\degr$ increases the obstacle distance from about 1.6\,$R_{\rm p}$ to 1.7\,$R_{\rm p}$.

\begin{figure}
\centering\includegraphics[width=\hsize,clip]{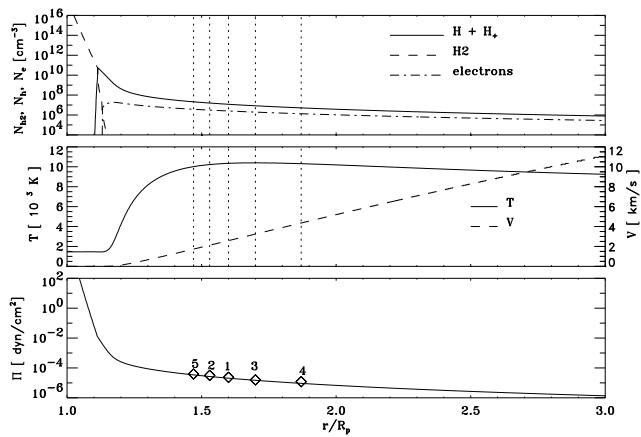}
\caption{Top: radial profiles of the planetary atmospheric number densities of molecular hydrogen (H$_2$; solid line), atomic hydrogen (neutral plus ions; dashed line), and electrons (dash-dotted line). Middle: temperature and velocity of the escaping atmospheric particles as a function of distance from the planetary center. Bottom: sum of the dynamic and thermal pressures ($\Pi$) in the atmospheric hydrodynamic flow as a function of the radial distance.  The diamonds and the corresponding vertical dotted lines indicate the position of the stagnation points for the stellar wind cases 1 to 5}.
\end{figure}

The pressure balance for the slow wind (Case 1) is reached at the distance of about 1.6\,$R_{\rm p}$. At this point the hydrogen number density $N_h$ (neutrals and ions) is equal to 1.0$\times$10$^{7}$\,cm$^{-3}$.  In case of an inclined magnetic field (Case 3; $\theta_{\rm B}=45\degr$), the minimal stand-off distance increases slightly to 1.7\,$R_{\rm p}$.

The pressure balance for the fast wind (Case 2) corresponds to a smaller distance of about 1.53\,$R_{\rm p}$, where the hydrogen number density is about 1.3$\times$10$^{7}$\,cm$^{-3}$. In Case 4, the pressure balance distance is located at about 1.9\,$R_{\rm p}$  with a total hydrogen number density of about
4.5$\times$10$^{6}$\,cm$^{-3}$. In Case 5, the pressure balance occurs at about 1.47\,$R_{\rm p}$, where the hydrogen number density is about 1.6$\times$10$^{7}$\,cm$^{-3}$. The pressure balance distances are much closer to the planet than those estimated in previous studies ($\ga$5\,$R_{\rm p}$, i.e. outside the Roche lobe; e.g. \citet[]{Khodachenko2015,Sha_2016}), which neglect the induced magnetic field.

In addition to the pressure balance, we consider also the penetration of the neutral atmospheric particles into the stellar wind. For these particles,
we take into account radiative ionization and charge exchange processes. The newly born ions are captured by the IMF and move away together with the magnetized stellar wind flow. From our simulations, we estimate that about $7.5\times 10^{9}$\,g\,s$^{-1}$ H atoms are ionized and removed by the stellar wind flow. The corresponding loss of neutral H atoms is about $3.5\times 10^{10}$\,g\,s$^{-1}$, which is in agreement with studies by \citet{Murray2009,Khodachenko2015,Sha_2016}. Although the H$^+$ loss rates are about 4.6 times lower than the thermal escape of neutral H atoms, one should still consider that the H$^+$ loss rate is about twice higher than that of suprathermal H atoms, which is about $3.5\times 10^{9}$\,g\,s$^{-1}$ \citep{Shematovich2010}.

We remark that for all  five stellar wind cases considered in our study we obtained obstacle boundaries closer to the planet than that yielding the best-fit to the Ly-$\alpha$ transit observations (2.9\,$R_{\rm p}$; KIS14). By comparing the results of the present study with the parameters that have been obtained by KIS14 to reproduce the observed Ly-$\alpha$ transit absorption, one can expect that our resulting planetary obstacles and the related stellar wind parameters most likely would not reproduce the observations. If a planetary obstacle with a stand-off distance like that assumed by KIS14 at about 3\,$R_{\rm p}$ is indeed necessary for fitting the Ly-$\alpha$ transit observations, then HD\,209458b should likely have a weak intrinsic magnetic field.

Table\,2 shows the estimated strength of the planetary magnetic moments $\mathcal{M}$ necessary to push the planetary obstacle to a distance of about 3\,$R_{\rm p}$ for the stellar wind cases 1 to 5. The effect of the intrinsic planetary magnetic field was estimated just by adding the planetary magnetic pressure term to the pressure balance equation.  Because the wind of HD\,209458 has likely a non-zero $B_w$, HD\,209458b most likely has an intrinsic magnetic moment with a strength of about 13--22\% that of Jupiter's. Our study also shows that accurate modeling of Ly-$\alpha$ transit observations should not neglect intrinsic and induced magnetic fields, as well as the plasma environment in the planet's vicinity.
\begin{table}
\caption{Planetary magnetic moments needed to push the obstacle to about 3\,$R_{\rm p}$, necessary for the reproduction of the HST Ly-$\alpha$ transit observations for the five stellar wind cases considered here.}
\begin{center}
\begin{tabular}{lcc}
\hline
Parameter cases & $\mathcal{M}$ [A m$^{-2}$] & $\mathcal{M}$ [$\mathcal{M}_{\rm Jup}$]  \\
\hline
Case 1: slow wind & $2.5\times 10^{26}$             & 0.16 \\
Case 2: fast wind & $3.0\times 10^{26}$             & 0.19 \\
Case 3: slow wind, inclined IMF & $2.0\times 10^{26}$& 0.13 \\
Case 4: KIS14     & $1.6\times 10^{26}$             & 0.1 \\
Case 5: KIS14 with B$_{\rm w}$ & $3.5\times 10^{26}$ & 0.22 \\
\hline
\end{tabular}
\end{center}
\end{table}
\section{CONCLUSION}\label{sec:conclusion}
We apply a 3D MHD model to the stellar wind flow around the planetary obstacle of the hot Jupiter HD\,209458b, in combination with a hydrodynamic upper atmosphere model.  We model the hydrodynamically expanding hydrogen atmosphere due to the absorption of the EUV flux from its host star. In this system, the EUV flux has a rather high intensity, which is more than 200 times larger than that for present Earth. Such EUV flux provides sufficient heating of the upper atmosphere to drive the hydrodynamic outflow of the hydrogen atoms. In addition to EUV heating, we account for dissociation, ionization, and recombination processes and focus only on the aspect of the interaction between the escaping neutral hydrogen atoms and the magnetized stellar wind plasma flow. In particular, we analyzed the effect of a strong enhancement of the stellar wind magnetic field in front of the planetary obstacle and its influence on the total pressure and position of the boundary.

Numerical solutions of the stellar wind interaction with an assumed non-magnetic HD\,209458b-like planet are obtained for five sets of stellar wind parameters. The results of the MHD simulations indicate that a strong magnetic field piles up in front of the planetary obstacle, where the magnetic pressure dominates the gas pressure of the stellar wind loaded by the ionized planetary particles.
An important feature is that the maximum of the total pressure at the stagnation point is much larger than the dynamic pressure of the undisturbed stellar wind. This is due to the strong influence of the induced magnetic field, which can move the planetary obstacle stand-off distance closer towards the planet, as compared to cases where the induced magnetic field is neglected.

This indicates that Ly-$\alpha$ transit observations can give important clues to understand how an exoplanet's upper atmosphere reacts to the stellar wind. Another important effect of the interaction between the stellar wind and the expanding planetary upper atmosphere is the pile-up of ions near the stagnation point due to charge exchange processes. By comparing our results of an assumed non-magnetic HD\,209458b-like exoplanet to the stellar wind parameters and planetary obstacle obtained by KIS14 from the fit to the observed Ly-$\alpha$ transit observations, we find that HD\,209458b should likely have an  intrinsic magnetic moment of about 13--22\% that of Jupiter's. This value is larger than that predicted by KIS14.

An inclination of the IMF relative to the stellar wind flow leads to a strong asymmetry of the flow structure, as well as of the magnetopause position. The total pressure maximum is shifted away from the $X$-axis, and the minimum distance between the planet and the magnetopause becomes a bit larger. Taking an inclination angle of $45\degr$, we estimate that an intrinsic planetary magnetic moment of about $2.0\times10^{26}$\,A\,m$^{-2}$ is necessary to shift the magnetopause to the distance of about $3 R_{\rm p}$, in order to agree with the Ly-$\alpha$ transit observations.

Finally, our results show that the atmospheric loss rate of a weakly magnetized HD\,209458b-like hot Jupiter is dominated by the EUV-driven hydrodynamic escape of H atoms, which is of the order of about $3.5\times 10^{10}$\,g\,s$^{-1}$, which is about 4.6 times and about 10 times larger than the loss rates of H$^+$ ions and suprathermal H atoms, respectively.

\section*{ACKNOWLEDGMENTS}
 The authors thank the anonymous referee for their useful comments. HL, PO and NVE acknowledge support from the Austrian Science Fund (FWF) project P25256-N27 ``Characterizing Stellar and Exoplanetary Environments via Modeling of Lyman-$\alpha$ Transit Observations of Hot Jupiters''. The authors acknowledge the support by the FWF NFN project S11601-N16 ``Pathways to Habitability: From Disks to Active Stars, Planets and Life'', and the related FWF NFN subprojects, S11604-N16 ``Radiation \& Wind Evolution from T Tauri Phase to ZAMS and Beyond'' (CJ), S11606-N16 ``Magnetospheric Electrodynamics of Exoplanets'' (MLK) and S11607-N16 ``Particle/Radiative Interactions with Upper Atmospheres of Planetary Bodies Under Extreme Stellar Conditions'' (KK, HL, NVE). DK, LF, and NVE acknowledge also the Austrian Forschungsf\"orderungsgesellschaft FFG project ``TAPAS4CHEOPS'' P853993. The authors further acknowledge support by the Russian Foundation of Basic Research grants No 15-05-00879-a (NVE, AVM) and No 16-52-14006 (NVE, AVM, IFS). MLK also acknowledges support by FWF projects I2939-N27, P25587-N27, P25640-N27, and the Leverhulme Trust Grant IN-2014-016.

\label{lastpage}


\begin{thebibliography}{}
	
\bibitem[\protect\citeauthoryear{Adams}{2011}]{Adams2011}
Adams, F. C., 2011, Astrophys. J., 730, 27

\bibitem[\protect\citeauthoryear{Ben-Jaffel}{2007}]{Be_2007}
Ben-Jaffel, L., 2007, Astrophys. J., 709, 1284

\bibitem[\protect\citeauthoryear{Bisikalo et al.}{2013}]{Bisikalo2013}
Bisikalo, D., Kaygorodov, P., Ionov, D., Shematovich, V., Lammer, H., Fossati, L., 2013,  Astrophys. J., 764, 19

\bibitem[\protect\citeauthoryear{Chadney et al.}{2015}]{Ch_2015}
Chadney, J. M., Galand, M., Unruh, Y. C., Koskinen, T. T., Sanz-Forcada, J., 2015, Icarus, 250, 357

\bibitem[\protect\citeauthoryear{Chadney et al.}{2016}]{Ch_2016}
Chadney, J. M., Galand, M., Koskinen, T. T., Miller, S., Sanz-Forcada, J., Unruh, Y. C., Yelle, R. V., 2016, Astron. Astrophys., 587, A87

\bibitem[\protect\citeauthoryear{Christie et al.}{2016}]{Christie2016}
Christie, D., Arras, P., Li, Z.-Y., 2016, Astrophys. J., 820, 3

\bibitem[\protect\citeauthoryear{Dong et al.}{2017}]{Dong2017}
 Dong, Ch., Lingam, M., Ma, Y., Cohen, O., 2017, Astrophys. J. Lett., 837, L26

\bibitem[\protect\citeauthoryear{Cohen et al.}{2011}]{Cohen2011}
 Cohen, O., Kashyap, V. L., Drake, J. J., Sokolov, I. V., Garraffo, C., Gombosi, T. I., 2011, Astrophys. J., 733, 67

\bibitem[\protect\citeauthoryear{Cohen et al.}{2014}]{Cohen2014}
 Cohen, O., Drake, J., Glocer, A., Garraffo, C., Poppenhaeger, K., Bell, J.M., Ridley, A.J., Gombosi, T.I., Astrophys. J., 790, 57

\bibitem[\protect\citeauthoryear{Cohen et al.}{2015}]{Cohen2015}
 Cohen, O., Ma, Y., Drake, J.J.,  Glocer, A.,  Garraffo, C.,  Bell, J. M., Gombosi, T. I., 2015, The Astrophys. J., 806, 41

\bibitem[\protect\citeauthoryear{Cook \& Metzger}{1964}]{Be_1964}
Cook, G. R., Metzger, P. H., 1964, J. Opt. Soc. Am., 54, 968

\bibitem[\protect\citeauthoryear{Cubillos et al.}{2017}]{Cub_2017}
Cubillos, P., Erkaev, N. V. Juvan, I., Fossati, L., Johnstone, C. P., Lammer, H., Lendl, M., Odert, P., Kislyakova, K. G., 2017, MNRAS, 466, 1868

\bibitem[\protect\citeauthoryear{Erkaev et al.}{2016}]{Erkaev2016}
Erkaev, N. V., Lammer, H., Odert, P., Kislyakova, K. G., Johnstone, C. P., G\"udel, M., Khodachenko, M. L., 2016, MNRAS, 460, 1300

\bibitem[\protect\citeauthoryear{Farrugia et al.}{2008}]{Far2008}
Farrugia, C. J., Erkaev, N. V., Taubenschuss, U. S., Vladimir A., Smith, C. W., Biernat, H. K., 2008, JGR, 113, A00B01

\bibitem[\protect\citeauthoryear{Farrugia et al.}{2009}]{Far2009}
Farrugia, C. J., Erkaev, N. V., Maynard, N. C., Richardson, I. G., Sandholt, P. E., Langmayr, D., Ogilvie, K. W., Szabo, A., Taubenschuss, U., Torbert, R. B., Biernat, H. K., 2009, Adv. Space Res., 44, 1288

\bibitem[\protect\citeauthoryear{Fossati et al.}{2017}]{fossati2017}
Fossati, L., Erkaev, N. V., Lammer, H., et al., Astron. Astrophys., 598, A90

\bibitem[\protect\citeauthoryear{Garcia Munoz}{2007}]{GarciaMunoz2007}
Garcia Munoz, A., 2007, Planet. Space. Sci., 55, 1426

\bibitem[\protect\citeauthoryear{Grie{\ss}meier et al.}{2004}]{Griessmeier2004}
Grie{\ss}meier, J.-M., Stadelmann, A., Penz, T., Lammer, H., Selsis, F., Ribas, I., Guinan, E. F., Motschmann, U., Biernat, H. K., Weiss, W. W., 2004, Astron. Astrophys., 425, 753

\bibitem[\protect\citeauthoryear{Grie{\ss}meier et al.}{2007}]{Griessmeier2007}
Grie{\ss}meier, J.-M., Zarka, P., Spreeuw, H., 2004, Astron. Astrophys., 475, 359

\bibitem[\protect\citeauthoryear{Guo}{2011}]{Guo2011}
Guo, J. H., 2011, Astrophys. J., 733, 98

\bibitem[Guo \& Ben-Jaffel(2016)]{guo2016}
Guo, J.~H., Ben-Jaffel, L., 2016, Astrophys. J., 818, 107

\bibitem[\protect\citeauthoryear{Janhunen}{2000}]{Janh2000}
Janhunen, P., 2000, J. Comp. Phys., 160, 649

\bibitem[\protect\citeauthoryear{Johnstone et al.}{2015}]{Johnstone2015}
Johnstone, C. P., G\"udel, M., L\"uftinger, T., Toth, G., Brott, I., 2015, Astron. Astrophys., 577, A27

\bibitem[\protect\citeauthoryear{Khodachenko et al.}{2015}]{Khodachenko2015}
Khodachenko, M. L., Shaikhislamov, I. F., Lammer, H., Prokopov, P. A., 2015, Astrophys. J., 813, 50

\bibitem[\protect\citeauthoryear{Kislyakova et al.}{2013}]{Kislyakova2013}
 Kislyakova, K. G., Lammer, H., Holmstr\"om, M., Panchenko, M., Odert, P., Erkaev, N. V., Leitzinger, M., Khodachenko, M. L., Kulikov, Yu. N., G\"udel, M., Hanslmeier, A., 2013, Astrobiology, 13, 1030

\bibitem[\protect\citeauthoryear{Kislyakova et al.}{2014a}]{Kislyakova2014a}
 Kislyakova, K. G., Johnstone, C. P., Odert, P., Erkaev N. V., Lammer, H., L\"uftinger, T., Holmstr\"om, M., Khodachenko, M. L., G\"udel, M., 2014a, Astron. Astrophys., 562, A116

\bibitem[\protect\citeauthoryear{Kislyakova et al.}{2014b}]{Kislyakova2014b}
Kislyakova, K. G., Holmstr\"om, M., Lammer, H., Odert, P., Khodachenko, M. L., 2014b,
Science, 346, 6212

\bibitem[\protect\citeauthoryear{Koskinen et al.}{2010}]{Koskinen2010}
Koskinen T. T., Cho J. Y., Achilleos N., Aylward A. D., 2010,
Astrophys. J., 722, 178

\bibitem[\protect\citeauthoryear{Koskinen et al.}{2013a}]{Koskinen2013a}
Koskinen T., Harris M., Yelle R., Lavvas P., 2013a, Icarus, 226,
1678

\bibitem[\protect\citeauthoryear{Koskinen et al.}{2013b}]{Koskinen2013b}
Koskinen T., Yelle R., Harris M., Lavvas P., 2013b, Icarus, 226,
1695

\bibitem[\protect\citeauthoryear{Lammer et al.}{2016}]{Lam2016}
Lammer, H., Erkaev, N. V., Fossati, L., Juvan, I., Odert, P., Guenther, E.,
Kislyakova, K. G., Johnstone, C., L\"uftinger, T., G\"{u}del, M., 2016, MNRAS, 461, L62

\bibitem[\protect\citeauthoryear{Lavvas et al.}{2014}]{Lavv2014}
Lavvas P., Koskinen T., Yelle R. V., 2014, Astrophys. J., 796, 15

\bibitem[\protect\citeauthoryear{Matsakos et al.}{2015}]{Matsakos2015}
Matsakos, T., Uribe, A., K\"onigl, A., 2015, Astron. Astrophys., 578, A6

\bibitem[\protect\citeauthoryear{Murray-Clay et al.}{2009}]{Murray2009}
Murray-Clay, R. A., Chiang, E. I., Murray, N., 2009, Astrophys. J., 693, 23

\bibitem[\protect\citeauthoryear{Owen \& Jackson}{2012}]{Owen2012}
Owen, J. E., Jackson, A. P., 2012, MNRAS, 425, 2931

\bibitem[\protect\citeauthoryear{Owen \& Adams}{2014}]{Owen2014}
Owen, J. E., Adams, F. C., 2014, MNRAS, 444, 3761

\bibitem[\protect\citeauthoryear{Powell}{1994}]{Powell1994}
Powell, K.G., 1994, ``An approximate Riemann solver for magnetohydrodynamics (that works in more than one dimension)'', ICASE Rep. 94-24, NASA Langley Res. Cent., Hampton, VA

\bibitem[\protect\citeauthoryear{Powell et al.}{1999}]{Powell1999}
Powell, K.G.,  Roe, P.L.,  Linde, T.J.,  Gombosi, T.I., De Zeeuw, D.L., 1999, J. Comp. Phys. 154, 28

\bibitem[\protect\citeauthoryear{Penz et al.}{2008}]{Penz2008}
Penz, T., Erkaev, N. V., Kulikov, Yu. N. Langmayr, D., Lammer, H., Micela, G., Cecchi-Pestellinig, C., Biernat, H. K., Selsis, F., Barge, P., Deleuil, M., L\'{e}ger, A., 2008, Planet Space Sci., 56, 1260

\bibitem[\protect\citeauthoryear{Salz et al.}{2016}]{Salz2016}
Salz M., Czesla S., Schneider P., Schmitt J., 2016, Astron. Astrophys., 586, A75

\bibitem[\protect\citeauthoryear{Shaikhislamov et al.}{2014}]{Sha_2014}
Shaikhislamov, I. F. , Khodachenko, M. L., Sasunov, Yu. L., Lammer, H., Kislyakova, K. G., Erkaev, N. V., 2014, Astrophys. J., 795, 132

\bibitem[\protect\citeauthoryear{Shaikhislamov et al.}{2016}]{Sha_2016}
Shaikhislamov, I. F., Khodachenko, M. L., Lammer, H., Kislyakova, K. G., Fossati, L., Johnstone, C. P., Prokopov, P. A., Berezutsky, A. G., Zakharov, Y. P., Posukh, V. G., 2016, Astrophys. J., 832, 173

\bibitem[\protect\citeauthoryear{Shematovich}{2010}]{Shematovich2010}
 Shematovich, V. I., 2010, Solar System Res., 44, 96

\bibitem[\protect\citeauthoryear{Shematovich et al.}{2014}]{She_2014}
Shematovich, V. I., Ionov, D. E., Lammer, H., 2014, Astron. Astrophys., 571, A94

\bibitem[\protect\citeauthoryear{Southworth}{2010}]{Southworth2010}
Southworth, J., 2010, MNRAS, 408, 1689

\bibitem[\protect\citeauthoryear{Stone \& Proga}{2009}]{Stone2009}
Stone, J. M., Proga, D., 2009, Astrophys. J., 694, 205

\bibitem[\protect\citeauthoryear{Tian et al.}{2005}]{Tian2005}
Tian, F., Toon, O. B., Pavlov, A. A., De Sterck, H., 2005, Astrophys. J., 621, 1049

\bibitem[\protect\citeauthoryear{Tilley et al.}{2016}]{Tilley2016}
Tilley, M. A., Harnett, E. M., Winglee, R. M., 2016, Astrophys. J., 827, 77

\bibitem[\protect\citeauthoryear{Trammell et al.}{2011}]{Trammell2011}
Trammell, G. B., Arras, P., Li, Z.-Y., 2011, Astrophys. J., 728, 152

\bibitem[\protect\citeauthoryear{Trammell et al.}{2014}]{Trammell2014}
Trammell, G. B., Li, Z.-Y., Arras, P., 2014, Astrophys. J., 788, 161

\bibitem[\protect\citeauthoryear{Tremblin \& Chiang}{2013}]{Tremblin2013}
Tremblin, P., Chiang, E., 2013, MNRAS, 428, 2565

\bibitem[\protect\citeauthoryear{Vidal-Madjar et al.}{2003}]{Vidal-Madjar2003}
Vidal-Madjar, A., Lecavelier Des Etangs, A., D\'esert, J.-M., Ballester,
G.E., Ferlet, R., H\'ebrard, G., Mayor, M., 2003, Nature, 422, 143

\bibitem[\protect\citeauthoryear{Weber et al.}{2017}]{Weber2017}
 Weber, C., Lammer, H., Chadney, J. M., Grie{\ss}meier, J.-M., Rucker, H. O.,
Vocks, C., Macher, W., Shaikhislamov, I. F., Khodachenko, M. L., Odert, P., Kislyakova, K. G., 2017, MNRAS, accepted

\bibitem[\protect\citeauthoryear{Welling \& Liemohn}{2016}]{Welling2016}
 Welling, D. T.,  Liemohn, M.W., 2016, J. Geophys. Res., 121, 5559

\bibitem[\protect\citeauthoryear{Yelle}{2004}]{Ye_2004}
Yelle, R.V., 2004, Icarus, 170, 167

\end{thebibliography}
\end{document}